\documentstyle[epsf,epsfig,aps]{revtex}
\draft
\input epsf
\begin{document}
\twocolumn[\hsize\textwidth\columnwidth\hsize\csname@twocolumnfalse%
\endcsname
\title{Induction of dc voltage, proportional to the persistent current, by
external ac current on system of inhomogeneous superconducting loops}
\author{S. V. Dubonos, \ V.  I.  Kuznetsov, \ I.  N. Zhilyaev, \ A. V.
Nikulov, and  A. A. Firsov}

\address{Institute of Microelectronics Technology and High Purity
Materials, Russian Academy of Sciences, 142432 Chernogolovka, Moscow
District, RUSSIA}

\maketitle
\begin{abstract}
{A dc voltage induced by an external ac current is observed in system of
asymmetric mesoscopic superconducting loops. The value and sign of this dc
voltage,  like the one of the persistent current, depend in a periodical
way on a magnetic field with period corresponded to the flux quantum within
the loop. The amplitude of the oscillations does not depend on the
frequency of the external ac current (in the investigated region $100 \ Hz$
- $1 \ MHz$) and depends on its amplitude. The latter dependence is not
monotonous. The observed phenomenon of rectification is interpreted as a
consequence of a dynamic resistive state induced by superposition of the
external current and the persistent current. It is shown that the dc
voltage can be added in system of loops connected in series: the dc voltage
oscillations with amplitude up to $10 \ \mu V$ were observed in single
loop, up to $40 \ \mu V$ in a system of 3 loops and  up to $300 \ \mu V$ in
a system of 20 loops.} \end{abstract}  ] \narrowtext

It is well known that the persistent current $j_{p} = 2en_{s}v_{s}$ is
 observed in superconductor \cite{tink75} because of the quantization of
 the momentum circulation $$\oint_{l}dl p =  \oint_{l}dl (mv + 2eA) =
m\oint_{l}dlv + 2e\Phi = n2\pi \hbar \eqno{(1)} $$ Its value and sign are
periodical function of the magnetic flux $\Phi$ inside a superconducting
loop with weak screening  $LI_{p} = Lsj_{p} < \Phi_{0}$ since the quantum
value of the velocity  circulation $$\oint_{l}dl v_{s} = \frac{2\pi
\hbar}{m} (n -\frac{\Phi}{\Phi_{0}}) \eqno{(2)} $$ should satisfy the
requirements of the energy minimum, i.e. the $n - \Phi/\Phi_{0}$ value
changes between -1/2 and 1/2 with $\Phi = BS + LI_{p}\simeq BS$ variation
\cite{tink75}. Here $B$ is the magnetic induction induced by an external
magnet;  $S = \pi r^{2}$ is the area of the loop.

The persistent current can exist permanently only in the closed
superconducting state, when the phase coherence is not broken along the
whole of  loop $l$ and its resistance $R_{l} = 0$. But according to the
Little-Parks  (LP) experiment \cite{little} and theory \cite{Kulik} the
persistent current $I_{p} \neq 0$ is observed not only in superconducting
state, i.e. at  $R_{l} = 0$, but also at  $R_{l} > 0$. It is also known
that a potential difference $V = (<\rho /s>_{l_{s}} - <\rho /s>_{l})l_{s}I$
should be observed on a segment $l_{s}$ of an conventional loop with
inhomogeneous resistivity $\rho$ or section $s$ if a current $I =
\oint_{l}E dl/R_{l}$ is induced by the Faraday's voltage $\oint_{l}E dl =
-(1/c)d\Phi/dt$ and the average resistance along the segment $R_{ls}/l_{s}
= <\rho/s>_{l_{s}} = \int_{l_{s}} dl \rho/sl_{s}$ differs from the one
along the loop $R_{l}/l = <\rho/s>_{l} = \oint_{l} dl \rho/sl$.

Therefore voltage oscillations $V(\Phi/\Phi_{0}) = (R_{ls}/l_{s} -
R_{l}/l)l_{s}I_{p}(\Phi/\Phi_{0}) \propto  (\overline{n} -\Phi/\Phi_{0})$
can be assumed at a segment of an inhomogeneous loop. Here $\overline{n}$
is  close, at $\Phi/\Phi_{0} \neq n + 1/2$, to an integer number $n$
corresponding to minimum $(n - \Phi/\Phi_{0})^{2}$ and $(\overline{n}
-\Phi/\Phi_{0}) = 0$ at $\Phi/\Phi_{0} = n + 1/2$. The persistent current
$I_{p} = 0$ at $\Phi/\Phi_{0} = n$ and $\Phi/\Phi_{0} = n + 1/2$
\cite{tink75}, has maximum values between $\Phi/\Phi_{0} = n$ and
$\Phi/\Phi_{0} = n + 1/2$ and has minimum values between $\Phi/\Phi_{0} = n
+ 1/2$ and $\Phi/\Phi_{0} = n + 1$ since the energy difference between
adjacent permitted states is much higher than $k_{B}T$ in any real
superconducting loop \cite{QuaForce} when the $\Phi/\Phi_{0}$ value is not
close to $ n + 1/2$. At $\Phi/\Phi_{0} = n + 1/2$ two permitted states ($n
- \Phi/\Phi_{0} = 1/2$ and $n - \Phi/\Phi_{0} = -1/2$) with opposite
velocity direction have the same minimum energy.

The $V(\Phi/\Phi_{0})$ oscillations, like the persistent current
oscillations $I_{p}(\Phi/\Phi_{0})$, were observed in our previous work
\cite{NANO2002} on asymmetric Al loop. We can observe these oscillations in
a narrow temperature region $0.988-0.994 T_{c}$, where $T_{c}$ corresponds
with the midpoint of the superconducting resistive transition. Its
amplitude increases in this region with temperature lowering. Such
temperature dependence is evidence of influence of an external electric
noise. In the present work the influence of an external ac current $I_{ac}
= \Delta I \sin(2\pi ft)$ (with different frequency $f$ and amplitude
$\Delta I$) on the $V(\Phi/\Phi_{0})$ oscillations is investigated  at
lower temperatures $0.95-0.98 T_{c}$ in order to clear up a question what
noise can induce the dc voltage  $V(\Phi/\Phi_{0})$ in \cite{NANO2002}.

Systems of 3 and 20 (see Fig.1) asymmetric Al loops with the critical
temperature $T_{c} \approx 1.3 K$, the sheet resistance $R_{\diamond }
\approx 0.5 \ \Omega /\diamond $ at 4.2 K and the resistance ratio $R(300
K)/R(4.2 K) \approx 2$ were used for investigations. We used system instead
of single loop used in \cite{NANO2002} in order to verify a possibility of
summation of the dc voltage in the system of identical asymmetric loops
connected in series.  All loops have the same diameter $2r = l/\pi = 4 \
\mu m$, Fig.1 and thickness 40 nm. One half of each loop $l_{n}$ is narrow
with the linewidth $w_{n} = 0.2 \ \mu m$ (the section $s_{n} = 0.008 \ \mu
m^{2}$) and the other half $l_{w}$ is wide $w_{w} = 0.4 \ \mu m$  ($s_{w} =
0.016 \ \mu m^{2}$). These Al microstructures are prepared using an
electron lithograph developed on the basis of a JEOL-840A electron scanning
microscop. An electron beam of the lithograph was controlled by a PC,
equipped with a software package for proximity effect correction "PROXY".
The exposition was made at 25 kV and 30 pA. The resist was developed in
MIBK: IPA = 1: 5, followed by the thermal deposition of a high-purity Al
film 60 nm and lift-off in acetone. The substrates are Si wafers.

The measurements are performed in a standard helium-4 cryostat allowing us
to vary the temperature down to 1.2 K. The applied perpendicular magnetic
field $B$ was produced by a superconducting coil. The external ac current
$I_{ac}$ with the frequency from $f = 100 \ \ Hz$ to $f = 1 \ \ MHz$ and the
amplitude up to $\Delta I = 50 \ \mu A$ flows between current contacts I - I,
Fig.1. The voltage was measured between different potential contacts V1-V8,
Fig.1. The dc voltage variations down to 0.05 $\mu V$ could be detected.

\begin{figure}[bhb] \vspace{0.1cm}\hspace{-1.5cm}
\vbox{\hfil\epsfig{figure= 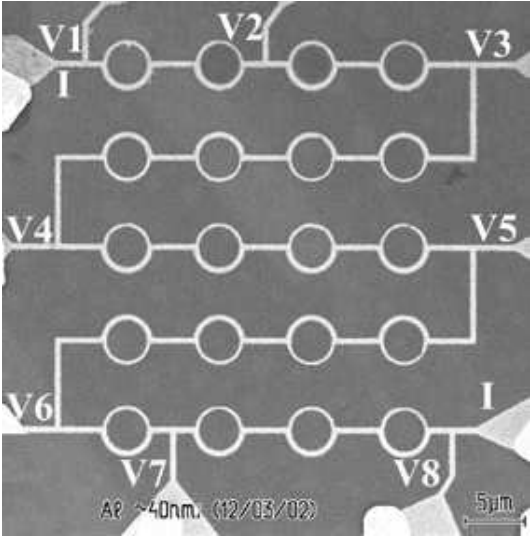,width=7cm,angle=0}\hfil}
\vspace{0.75cm} \caption{Electron micrograph of the system of 20 asymmetric
aluminum loops.  $I$ are the current contacts and $V1$,  $V2$, $V3$, $V4$,
$V5$, $V6$, $V7$, $V8$, are potential contacts.} \label{fig-1} \end{figure}

Our measurements showed that the external as current $I_{ac}$ with any
frequency from $f = 100 \ Hz$ to $f = 1 \ MHz$ induces the dc voltage
oscillations $V(\Phi/\Phi_{0})$ Fig.2 when its amplitude $\Delta I$ exceeds
a critical value $\Delta I_{cr}$ which decreases, like the superconducting
critical current $I_{c}$, with temperature and magnetic field increase.
Because of the latter the oscillations are observed only at high
$\Phi/\Phi_{0}$ values when the $I_{ac}$ amplitude is relatively low, for
example at $\Delta I = 3 \ \mu A$ on Fig.2. The oscillations
$V(\Phi/\Phi_{0})$ fade at higher $\Phi/\Phi_{0}$ (like the LP
oscillations, see for example \cite{repeat}) because of the suppression of
the superconductivity, i.e. $I_{p}$, in the wire defining the loops be the
magnetic field.

The amplitude of the oscillations $V(\Phi/\Phi_{0})$ does not depend on the
frequency of the external ac current (in the investigated region $100 \ Hz$
- $1 \ MHz$) and depends on its amplitude $\Delta I$. The latter dependence
is not monotonous Fig.2. The amplitude  $|V|_{max}$ observed at
$\Phi/\Phi_{0} \approx \pm 0.25$ mounts a maximum value at $\Delta I$
slightly  higher $\Delta I_{cr}$ and decreases further with the $\Delta I$
increase, Fig.2. The dependence $|V|_{max}(\Delta I)$ is close to
$|V|_{max} \propto 1/\Delta I$ at high $\Delta I$ values. The $\Delta I$
value corresponded to the maximum of  $|V|_{max}(\Delta I)$ and $\Delta
I_{cr}$ decreases with drawing $T$ near $T_{c}$ like the superconducting
critical current.

Our investigation has shown that the dc voltage can be easy summed up in the
system of identical asymmetric loops connected in series. The
$V(\Phi/\Phi_{0})$ oscillations with amplitude up to $ \approx 10 \ \mu V$
were observe on single loop, up to $\approx 40 \ \mu V$ on system of 3
loops and up to $\approx 300 \ \mu V$ on system of 20 loops, Fig.3.

The effect observed in our work may be interpreted as a consequence of a
transference of the loop with $I_{p} \neq 0$ to the resistive state by the
external current $I_{ac}$ or as a rectification of the ac current $I_{ac}$
because of an asymmetry of the current-voltage curves, Fig.4. This
asymmetry takes place and its value and sign are periodical function of the
magnetic flux $\Phi$ because of superposition of the external $I_{ac}$
current and the intrinsic persistent $I_{p}$ current.

\begin{figure}[bhb] \vspace{0.1cm}\hspace{-1.5cm}
\vbox{\hfil\epsfig{figure= 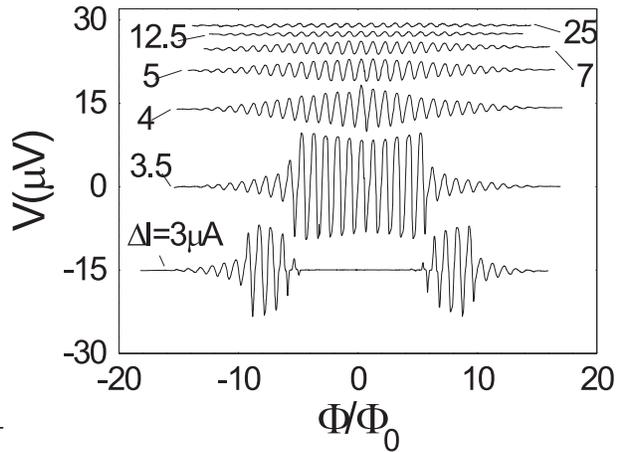,width=8cm,angle=0}\hfil}
\vspace{0.75cm} \caption{Oscillations of the dc voltage induced on single
loop by the ac current with the frequency $f = 2.03 kHz$ and amplitudes
$\Delta I = 3; 3.5; 4; 5; 7; 12.5; 25 \ \mu A$. $T = 1.280  \ K$ is
corresponding to $0.97T_{c}$. Except  $\Delta I =  3.5 \ \mu A$, all other
dependencies are displaced in the vertical direction. } \label{fig-1}
\end{figure}

The distribution of the external current $I_{ac} = I_{w} + I_{n}$ between
loop halves $l_{w}$ and $l_{n}$ is determined in the closed superconducting
state by the quantization of the velocity circulation (2):
$l_{n}\overline{v_{n}} - l_{w} \overline{v_{w}} = l_{n}j_{n}/2en_{sn} -
l_{w}j_{w}/2en_{sw} = l_{n}I_{n}/s_{n}2en_{sn} - l_{w}I_{w}/s_{w}2en_{sw} =
(2\pi \hbar /m) (\overline{n} - \Phi/\Phi_{0})$. The current density $j_{n}
= j_{w} = j_{ac} = I_{ac}/(s_{n}+s_{w})$ and the critical value $j_{c}$ is
achieved simultaneously (at $I_{ac} = (s_{n}+s_{w})j_{c}$)  in the both
halves at $\overline{n} - \Phi/\Phi_{0} = 0$, i.e. $I_{p} = 0$, since the
length $l_{n}$ equals $l_{w}$ and the densities of superconducting pairs
$n_{sn}$ and $n_{sw}$ are the same in the both halves.

At $\overline{n} - \Phi/\Phi_{0} \neq 0$ the densities $j_{n}$ and $j_{w}$
are different (for example $j_{n} = I_{ac}/(s_{n}+s_{w}) + I_{p}/s_{n} $
and $j_{w} = I_{ac}/(s_{n}+s_{w}) - I_{p}/s_{w}$, Fig.4, when the $I_{ac}$
direction coincides with the  $I_{p}$ one on the narrow half $l_{n}$) and
the persistent current decreases the critical $I_{ac}$ value. These
critical values in opposite directions  $|I_{ac}|_{c+} =
(s_{n}+s_{w})(j_{c} - I_{p}/s_{n})$ and $|I_{ac}|_{c-} =
(s_{n}+s_{w})(j_{c} - I_{p}/s_{w})$ are different at $I_{p} \neq 0$ since
$s_{w} > s_{n}$.

Since $|I_{ac}|_{c-} - |I_{ac}|_{c+} = (s_{n}+s_{w})(j_{c} - I_{p}/s_{n}) -
(s_{n}+s_{w})(j_{c} - I_{p}/s_{w}) = (s_{w}/s_{n} - s_{n}/s_{w})I_{p}
\approx 1.5I_{p}$ the $I_{p}$ and $(s_{n}+s_{w})j_{c}$ values can be
evaluated from measured $|I_{ac}|_{c-}$ and $|I_{ac}|_{c+}$ values. For
example it is followed from the values $|I_{ac}|_{c+} \approx 0.45 \ \mu A$
and $|I_{ac}|_{c-} \approx 1.0 \ \mu A$ (see the current-voltage curve on
Fig.4) measured at $\overline{n} - \Phi/\Phi_{0} \neq 0$ and $T \approx
0.99 T_{c}$ that $I_{p} \approx  (|I_{ac}|_{c-} - |I_{ac}|_{c+}) \approx
0.37 \ \mu A$ and $(s_{n}+s_{w})j_{c} \approx 1.6 \ \mu A$. Note should be
taken that the current-voltage curve Fig.4 at $|I_{ac}| >
(s_{n}+s_{w})j_{c}$ coincides with the one in the normal state and a
peculiarity is observed at $|I_{ac}| \approx  (s_{n}+s_{w})j_{c}$.

\begin{figure}[bhb] \vspace{0.1cm}\hspace{-1.5cm}
\vbox{\hfil\epsfig{figure= 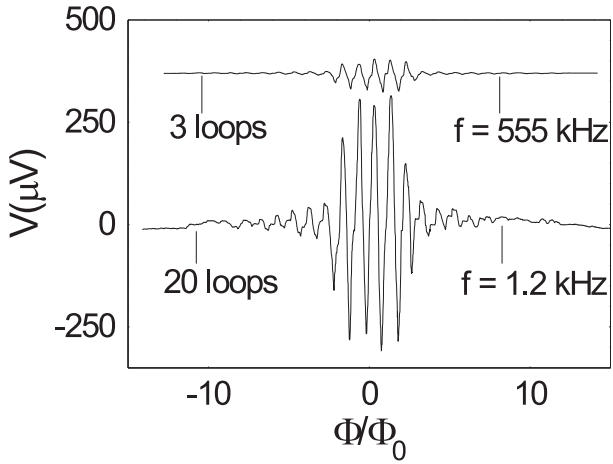,width=8cm,angle=0}\hfil}
\vspace{0.75cm} \caption{Oscillations of dc voltage induced on the system
of 20 loops by ac current with $f = 1.2 kHz$ and $\Delta I = 3.2 \ \mu A$
at $T = 1.245  \ K = 0.97T_{c}$ and on the system of 3 loops by $I_{ac}$
with $f = 555 kHz$ and $\Delta I = 4.5 \ \mu A$ at $T = 1.264 \ K =
0.96T_{c}$. Second dependence is displaced on 370 $\mu m$ in the vertical
direction. } \label{fig-1} \end{figure}

Since $(s_{n}+s_{w})j_{c} > |I_{ac}|_{c-} > |I_{ac}|_{c+}$ the voltage (the
resistive state) appears first at $\overline{n} - \Phi/\Phi_{0} \neq 0$ and
$|I_{ac}|_{c-} > \Delta I > |I_{ac}|_{c+}$ in only direction of the
external ac current $I_{ac} = \Delta I \sin(2\pi ft)$, see Fig.5. The
absence of the voltage at $\Delta I = 9 \ \mu A$ (corresponded to the
maximum of the $|V|_{max}(\Delta I)$ dependence) and $\Phi = 0$ (see the
upper oscillograph curve on Fig.5) means that $j_{c} > 9 \ \mu A/s_{w}
\approx 9 \ \mu A/0.016 \ \mu m^{2} \approx 5.6 \ 10^{8} \ A/m^{2}$ at $T
\approx 0.96T_{c}$ since the width of the wire between the loops and
contacts equals $s_{w} < s_{w} + s_{n}$, see Fig.1 and Fig.4. At the same
temperature and $\Delta I$ value but $\Phi = 0.75 \Phi_{0}$ the voltage is
observed in only direction of $I_{ac} = \Delta I \sin(2\pi ft)$,  (see the
second oscillograph curve on Fig.5). This means that $|I_{ac}|_{c-} >
\Delta I =9 \ \mu A > |I_{ac}|_{c+}$ at $T \approx 0.96T_{c}$ and $\Phi =
0.75 \Phi_{0}$ and consequently $I_{p} = s_{n}(j_{c} -
|I_{ac}|_{c+}/(s_{n}+s_{w})) > s_{n}(j_{c} - 9 \ \mu A/(s_{n}+s_{w})) > 9 \ \mu
A s_{n}^{2}/s_{w}(s_{n}+s_{w}) \approx 1.5 \ \mu A$. At $\Phi = 4.25
\Phi_{0}$  (see the third oscillograph curve on Fig.5) the steady voltage
is observed in the opposite (relatively the $\Phi = 0.75 \Phi_{0}$ case)
$I_{ac}$ direction and the unsteady voltage is observed in the other
direction. The latter takes place because of the $j_{c}$ decrease down to
$|I_{ac}|_{c-} = (s_{n}+s_{w})(j_{c} - I_{p}/s_{w}) \approx 9 \ \mu A$ in a
high magnetic field. At  $\Delta I > |I_{ac}|_{c-}$ the steady voltage is
observed in the both $I_{ac}$ directions (see the lower oscillograph curve
on Fig.5).

\begin{figure}[bhb] \vspace{0.1cm}\hspace{-1.5cm}
\vbox{\hfil\epsfig{figure=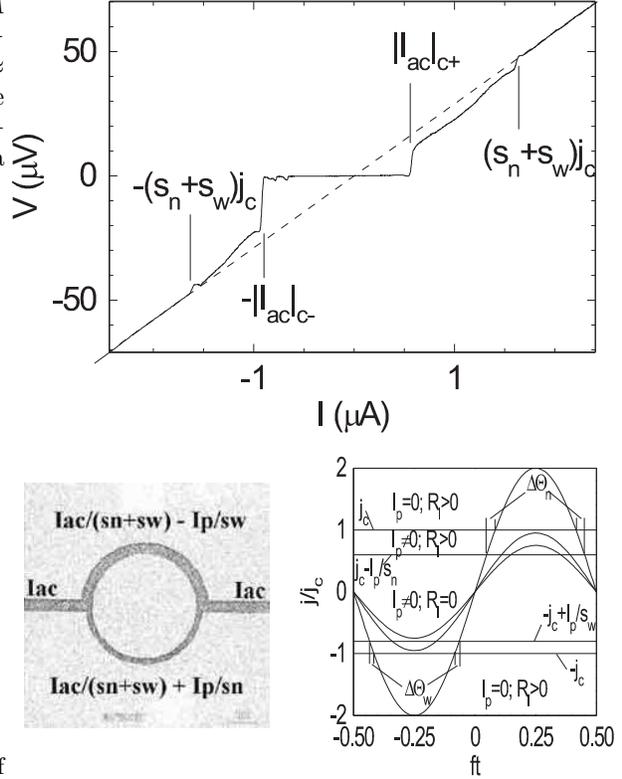,width=8cm,angle=0}\hfil}
\vspace{0.75cm} \caption{The current-voltage curve measured on a system of
2 loops at $T \approx 0.99T_{c}$, $\Phi \neq n\Phi_{0}$ and $\Phi \neq
(n+0.5)\Phi_{0}$ is shown in the above part. The change of the current
distribution between loop halves $l_{w}$ and $l_{n}$ because of the
persistent current $I_{p} \neq 0$ is indicated in the lower left side. The
drawing in the lower right side elucidates why the $|V|_{max}(\Delta I)$
dependence (see Fig.2 and 6) is not monotonous.} \label{fig-1} \end{figure}

The external current $|I_{ac}| <  (s_{n}+s_{w})j_{c}$ can not destroy
superconductivity in the both ($l_{w}$ and $l_{n}$) halves of the loop.
Therefore the resistive state observed at $|I_{ac}|_{c+} < |I_{ac}| <
(s_{n}+s_{w})j_{c}$ (and at $|I_{ac}|_{c-} < |I_{ac}| <
(s_{n}+s_{w})j_{c}$, see Fig.4) can not be static. When the $l_{n}$ (for
example) half transfers in the resistive state with $R_{n} > 0$ and $V(t) =
R_{n}I_{n} \neq 0$ at $I_{n} = I_{ac}s_{n}/ (s_{n}+s_{w}) + I_{p} >
s_{n}j_{c}$ the $I_{n}$ value decreases (because of the $I_{p}$ reduction
and the $I_{w}$ increase according to the law $L_{w}dI_{w}/dt = V(t)$) and
the $l_{n}$ half should return at $I_{n} < s_{n}j_{c}$ to the
superconducting state with $I_{n} = I_{ac}s_{n}/ (s_{n}+s_{w}) + I_{p} >
s_{n}j_{c}$. Thus, the superconducting state with both connectivity (i.e.
with $R_{n} > 0$ and $R_{n} = 0$) can not be stable and the loop should
switch between they with an intrinsic frequency $\omega $ at $|I_{ac}|_{c+}
< |I_{ac}| <  (s_{n}+s_{w})j_{c}$ (or at $|I_{ac}|_{c-} < |I_{ac}| <
(s_{n}+s_{w})j_{c}$).

\begin{figure}[bhb] \vspace{0.1cm}\hspace{-1.5cm}
\vbox{\hfil\epsfig{figure= 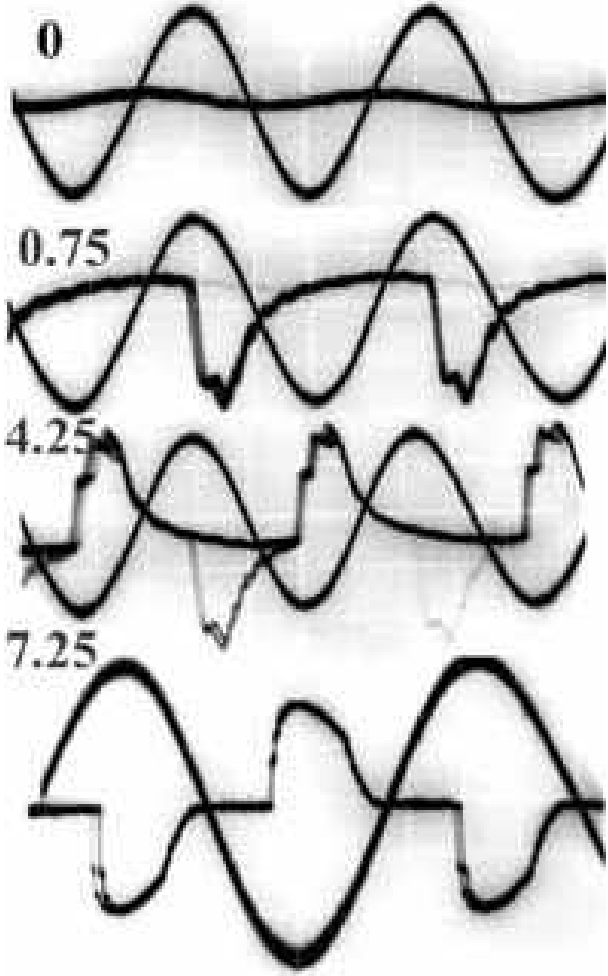,width=8cm,angle=0}\hfil}
\vspace{0.75cm} \caption{The oscillograph curves of the voltage measured on
a system of  3 loops and of the ac current inducing it at: $\Phi = 0$,
$\Phi = 0.75\Phi_{0}$, $\Phi = 4.25\Phi_{0}$ ($f = 11.53 \ kHz$, $\Delta I
= 9 \ \mu A$, $T \approx  0.96T_{c}$) and $\Phi = 7.25\Phi_{0}$ ($f = 1.53
\ kHz$, $\Delta I = 5.8 \ \mu A$, $T \approx  0.97T_{c}$).}  \label{fig-1}
\end{figure}

It should be noted that the half in which  $I_{ac}$ and $ I_{p}$ have
opposite directions remains all time in superconducting state in this
dynamic resistive state although the average in time voltage $V_{dc} =
\overline{V(t)}^{t}$  (measured, for example, on the current-voltage curve
Fig.4) is not equal zero. It is possible because of the switching between
superconducting state with different connectivity. The voltage would not be
observed at $|I_{ac}|_{c+} < |I_{ac}| <  s_{w}j_{c}$ (for example on  the
second oscillograph curve of Fig.5) if the $l_{n}$ half could remain
constantly in the norman state with $R_{n} > 0$ when superconducting state
in the loop is unclosed and $I_{p} = 0$.

The dc voltage $V_{dc} \neq 0$ can be observed on the superconducting half
($l_{w}$ for example) since the velocity $v_{s}$ and the momentum $p =
mv_{s} + 2eA$ of superconducting pairs should revert to the initial quantum
values, after the acceleration $dp/dt =2eE(t) = 2eV(t)/l_{w} = 2eR_{n}I{n}$
at $R_{n} > 0$, when the loop reverts to the closed superconducting state.
The same reason explains the observation of the persistent current at $R >
0$ in the LP experiment  \cite{QuaForce}. Since the $p$ change because of
$V(t)$ equals the average change  $\Delta p$ because of the quantization
$$V_{dc} = \overline{V(t)}^{t} = \overline{R_{n}I_{n}}^{t} = \Delta p\omega
l_{w}/2e \eqno{(3)} $$

The $\Delta p$ value is connected with the deviation of $I_{w}$ (and
consequently $I_{n} = I_{ac} - I_{w}$) from the quantum value $I_{w} =
I_{ac}s_{w}/(s_{n}+s_{w}) - I_{p}$ during a time $t_{R>0}$ when the
superconducting state is unclosed, i.e. $R_{n} > 0$. This change of the
current distribution  $\Delta I_{w} < I_{ac}/3 + I_{p}$ and $\Delta I_{w} <
3I_{p}$ since $I_{n} > 0$ and $I_{w} < s_{w}j_{c}$ in the dynamic resistive
state. The momentum circulation of superconducting pair changes from the
quantum value $2\pi \hbar n$ to $2e\Phi$ when the persistent current
changes from the quantum value $I_{p}$ to zero \cite{QuaForce}. Therefore
$\Delta p = \oint_{l}dl \Delta p/l < 3(2\pi \hbar /l)(\overline{n} -
\Phi/\Phi_{0})$ since $\Delta I_{w}< 3I_{p}$ and $2\pi \hbar \overline{n} -
2e\Phi = 2\pi \hbar (\overline{n} - \Phi/\Phi_{0})$. According to this
inequality and (3) the switching frequency $\omega = 2eV_{dc}/\Delta pl_{w}
> 1/3(e\pi \hbar )(l/l_{w}) V_{dc}/ (\overline{n} - \Phi/\Phi_{0}) \approx
1/3(0.4836 \ GHz/\mu V)V_{dc}/ (\overline{n} - \Phi/\Phi_{0})$ should
exceed 90 GHz at $V_{dc} \approx  70 \ \mu V$ observed on single loop at $|n
- \Phi/\Phi_{0}| \approx 1/4$ (see Fig.5 for example).

The change of the current distribution  $\Delta I_{w}$ determined by the
relaxation law $L_{w}dI_{w}/dt = -L_{w}dI_{n}/dt  = V(t) = R_{n}I_{n}$ (and
consequently $\Delta p$) is smaller than its maximum possible value if the
time of the relaxation to the equilibrium superconducting state $t_{R>0}$
during which $R_{n} > 0$ is shorter than the time $L_{w}/R_{n}$ of current
relaxation. The kinetic inductance $L_{w}$ exceeds the geometric one
$L_{g}$, $L_{w}/L_{g} \approx \lambda_{L}^{2}/s_{w} \gg 1$,  in the case of
weak screening,   $s_{w} \ll  \lambda _{L}^{2}$. The section of the loop
$s_{w} \approx 0.016 \ \mu m^{2}$ is smaller than the square of the London
penetration depth $\lambda_{L}^{2}(T)$ of Al at $T = 0.95-0.98T_{c}$,
$\lambda_{L}^{2}(T) =\lambda_{L}^{2}(0)/(1-T/T_{c}) \approx (0.05 \ \mu
m)^{2}/(1-T/T_{c}) \approx 0.1 \ \mu m^{2}$ at $1-T/T_{c} = 0.025$.
Therefore the inductance of $l_{w}$, $L_{w} = L_{kin} + L_{g} \approx
L_{kin} \approx \mu_{0}l_{w}\lambda_{L}^{2}(T)/s_{w} \approx 4 \ 10^{-11} \
G$ at $T = 0.975T_{c}$ and $L_{w}/R_{n} > L_{w}/R_{n,n} \approx 3 \
10^{-12} \ s$ since the resistance of $l_{n}$ in the normal state $R_{n,n}
\approx 15 \ \Omega $.

The observed value of the dc voltage $V_{dc} = \overline{V(t)}^{t} =
\overline{R_{n}I_{n}}^{t} < R_{n,n} \overline{I_{n}}^{t}$ is evidence of
weak variation of the current distribution since the value the average
current $\overline{I_{n}}^{t} > V_{dc}/R_{n,n}$ is close to the value
$I_{n} = I_{ac}s_{n}/(s_{n}+s_{w}) + I_{p}$ in the closed superconducting
state. For example, the value $V_{dc} \approx 7.4 \ \mu V$ measured at
$I_{ac} \approx 0.7 \ \mu A$ on single loop, see the current-voltage curve
on Fig.4, gives $\overline{I_{n}}^{t} > V_{dc}/R_{n,n} \approx 0.49 \ \mu
A$ whereas $ I_{ac}s_{n}/(s_{n}+s_{w})  \approx 0.23 \ \mu A$ and
$I_{ac}s_{n}/(s_{n}+s_{w}) + I_{p} \approx 0.6 \ \mu A$. The weak variation
of the current distribution means that $t_{R>0} < L_{w}/R_{n} \geq  3 \
10^{-12} \ s$.

Thus, the dc voltage oscillations $V(\Phi/\Phi_{0})$ like the persistent
current oscillations $I_{p}(\Phi/\Phi_{0})$ are observed since both
$\overline{I_{p}}^{t} \neq 0$ and $\overline{R_{l}}^{t} > 0$ in the dynamic
resistive state because of switching between superconducting states with
different connectivity. Only this state makes a contribution to
$V(\Phi/\Phi_{0})$ since $R_{l} = 0$ at $|I_{as}| < j_{c}(s_{n}+s_{w}) -
I_{p}(s_{n}+s_{w})/s_{n}$ and at $|I_{as}| > j_{c}(s_{n}+s_{w})$, $I_{p} =
0$ and the current-voltage curve does not differ from the one in the normal
state, see Fig.4.

Therefore the $V(\Phi/\Phi_{0})$ value can be evaluated by the relation $V
\approx (V_{dc,n}\Delta \Theta_{n} - V_{dc,w}\Delta \Theta_{w})/\Theta $,
where $\Delta \Theta_{n}$ and $\Delta \Theta_{w}$ are parts of the time
period $\Theta = 1/f$ during which $j_{c}(s_{n}+s_{w}) -
I_{p}(s_{n}+s_{w})/s_{n} < \Delta I | \sin(2\pi ft)| < j_{c}(s_{n}+s_{w})$
and $j_{c}(s_{n}+s_{w}) - I_{p}(s_{n}+s_{w})/s_{w} < \Delta I | \sin(2\pi
ft)| < j_{c}(s_{n}+s_{w})$ (see Fig.4). $V_{dc,n}$ and $V_{dc,w}$ are
average voltage during $\Delta \Theta_{n}$ and $\Delta \Theta_{w}$.

\begin{figure}[bhb] \vspace{0.1cm}\hspace{-1.5cm}
\vbox{\hfil\epsfig{figure= 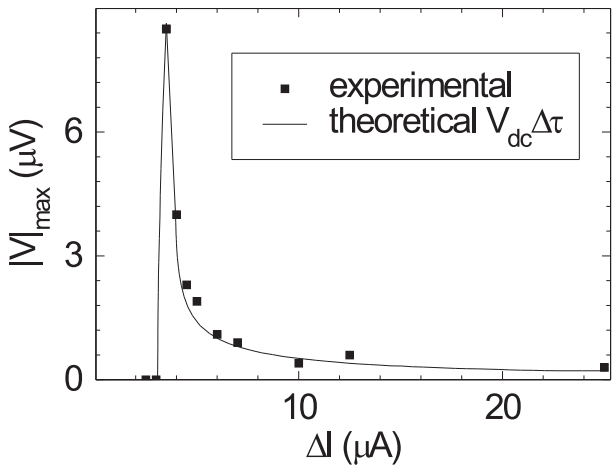,width=8cm,angle=0}\hfil}
\vspace{0.75cm} \caption{The dependence of the amplitude of the dc voltage
oscillations $V(\Phi/\Phi_{0})$ on the amplitude  $\Delta I$ of the
external ac current with the frequency  $f = 2.03 \ kHz$ measured on single
loop at $T = 1.280 \ K \approx  0.97T_{c}$ and $\Phi/\Phi_{0} \approx \pm
0.25$. The line is the theoretical dependence $V_{dc}(\Delta \Theta_{n} -
\Delta \Theta_{w})/\Theta = V_{dc} \Delta \tau (\Delta I) $ calculated at
$V_{dc} \approx 25 \ \mu V$, $I_{c} \approx 4.5 \ \mu A$ and  $I_{p}
\approx 0.5 \ \mu A$} \label{fig-1} \end{figure}

The dependence of $(\Delta \Theta_{n} - \Delta \Theta_{w})/\Theta = \Delta
\tau $ on the amplitude $\Delta I$ of the ac current, $\tau (\Delta I)$, is
not monotonous like the observed $|V|_{max}(\Delta I)$ dependence, Fig.6.
The $\tau (\Delta I)$ and $|V|_{max}(\Delta I)$ have maximum values at
$j_{c} - I_{p}/s_{n} < \Delta I/(s_{n}+s_{w})  < j_{c} - I_{p}/s_{w}$ when
$\Delta \Theta_{w} = 0$, see Fig.4, and the voltage is observed in only
direction of $I_{ac}$, see Fig.5. When $\Delta I/(s_{n}+s_{w})$ exceeds
$j_{c} - I_{p}/s_{w}$ the $\tau$ and $|V|_{max}$ values decrease sharply
but remain non-zero since  $\Delta \Theta_{n} > \Delta \Theta_{w}$, Fig.4,
because of the loop asymmetry. The following diminution of $|V|_{max}$ with
the $\Delta I$ increase, Fig.6, may be explained by the $\Delta \Theta_{n}$
and $\Delta \Theta_{w}$ diminution, Fig.4. The approximation
$|V|_{max}(\Delta I) = V_{dc}\tau (\Delta I)$, where $V_{dc} \approx
V_{dc,n} \approx V_{dc,w} \approx 25 \ \mu V$ and the $\tau (\Delta I)$
dependence is calculated at $I_{c} \approx 4.5 \ \mu A$ and  $I_{p} \approx
0.5 \ \mu A$, describes enough well the experimental $|V|_{max}(\Delta I)$
dependence observed at $T = 1.280 \ K \approx  0.97T_{c}$, Fig.6.

In conclusion, we have shown that the dc voltage oscillations observed in
\cite{NANO2002} are connected with a rectification of the ac current
because of an asymmetry of the current-voltage curves of asymmetric
superconducting loop at non-zero persistent current. Since the critical
amplitude of the ac current inducing the dc voltage oscillations decreases
down to zero with drawing $T$ near $T_{c}$, like the superconducting
critical current, any how weak electric noise, right down to intrinsic
noise, can induce these oscillations when the temperature is enough close
to $T_{c}$. Therefore the system of asymmetric superconducting loops can be
used as noise detector with maximum sensitivity. This system can be used
also as a source of high-frequency radiation if the switching of loops
between superconducting states with different connectivity are synchronous
in it.

We acknowledge useful discussions with V.A.Tulin. This work was financially
supported by the Presidium of Russian Academy of Sciences in the Program
"Low-Dimensional Quantum Structures".

\end{document}